\documentclass[twocolumn,showpacs,floatfix,superscriptaddress,amsmath,amssymb,pre]{revtex4}
\usepackage{mathrsfs}
\usepackage{txfonts}
\usepackage{amssymb}
\usepackage{graphicx}
\usepackage{hyperref}
\usepackage{ulem}
 \usepackage{overpic}

\usepackage{color}
 \usepackage{psfrag}
\usepackage{tabularx}
\usepackage{array}
\usepackage{placeins}
\makeatletter

\begin{document}

\title{Universal scaling relationship between classical and quantum correlations in critical quantum spin chains}

\author{Yan-Wei Dai}
\affiliation{State Key Laboratory of Optoelectronic Material and Technologies,
School of Physics, Sun Yat-sen University, Guangzhou 510275, China }
\affiliation{Centre for Modern Physics,
Chongqing University, Chongqing 400044, China}

\author{Xi-Hao Chen}
\affiliation{College of Materials Science and Engineering, Chongqing University, Chongqing 400044,China}
\affiliation{Centre for Modern Physics,
Chongqing University, Chongqing 400044, China}

\author{Sam Young Cho}
\altaffiliation{E-mail: sycho@cqu.edu.cn}
\affiliation{Centre for Modern
Physics, Chongqing University, Chongqing 400044, China}
\affiliation{Department of
Physics, Chongqing University, Chongqing 400044, China}

\author{Huan-Qiang Zhou}
\affiliation{Centre for Modern Physics,
Chongqing University, Chongqing 400044, China}
\affiliation{Department of
Physics, Chongqing University, Chongqing 400044, China}

\author{Dao-Xin Yao}
\affiliation{State Key Laboratory of Optoelectronic Material and Technologies,
School of Physics, Sun Yat-sen University, Guangzhou 510275, China }

\begin{abstract}
 We numerically investigate classical and quantum correlations
 in one-dimensional quantum critical systems.
 The infinite matrix product state (iMPS) representation is employed
 in order to consider an infinite-size spin chain.
 By using the infinite time-evolving block decimation algorithm,
 iMPS ground state wave functions are obtained at critical points
 for the transverse-field spin-$1/2$ XY model.
 From the ground state wave functions, we calculate classical and quantum correlations and mutual information. All of the correlations are found to exhibit a power-law decay with the increments
 of the lattice distance for both the transition lines
 of the Ising universality class and the Gaussian universality class.
 Such power-law scaling behaviors of the correlations manifest the existence of diversing correlation lengths, which means scale invariance. The critical features of the correlations can be characterized by introducing a critical exponent of the power-law decaying correlations.
 Similar to the critical exponent $\eta$ of the spin-spin correlation
 for the universality classes in the transverse-field XY model,
 we calculate the critical exponents of the two-spin classical and quantum correlations as well as that of the corresponding mutual information.
 All of the correlations have the same critical exponents, i.e., $\eta^{I}=\eta^{C}=\eta^{D}$ at a critical point, where the superscripts $I$, $C$, and $D$ stand
 for mutual information, classical correlation, and quantum correlation, respectively.
 Furthermore, the critical exponent $\eta$ of the spin-spin correlation
 is shown to relate to $\eta = \eta^\alpha/2$ with $\alpha \in \{ I, C, D\}$.
\end{abstract}
\pacs{03.67.Mn, 75.10.Pq, 75.40.Cx}

\maketitle

\section{Introduction}

 Quantum many-body systems have revealed many intriguing features and phenomena completely different from
 those of simple aggregations of individual particles.
 Of particular interest in such features are quantum phase transitions \cite{Sachdev,2}
 that are abrupt changes of groundstate wavefunction structure for varying system parameters at zero temperature.
 Such a structure change of groundstate wavefunction is driven by quantum fluctuations
 originating from the Heisenberg uncertainty principle.
 Conventionally, quantum phases of many-body systems can be characterized by using
 the long-distance behaviors, i.e., the scaling behaviors of spatial correlation functions \cite{Sachdev,2}.
 In contrast to the exponential decay of spatial correlation function in gapped phase systems,
 critical systems has a characteristic feature of spatial correlation function that
 is its power-law decay.
 The long-range properties of spatial correlations are the
 same for different materials in the same universality class.
 Thus a simple model can describe complicated real materials
 in the sense of the notion of universality.
 The critical exponents of power-law decays in the spatial correlations
 can then classify the universality classes of critical systems.

 In the past decades, from the perspective of quantum information,
 quantum entanglement-based measures have been introduced to characterize quantum phases \cite{3}.
 Much efforts have been made to developing
 and understanding quantitative measures of entanglement \cite{5,6,7,8}.
 Recently, remarkable progresses
 have been made particularly in more fundamental investigations of quantum criticality.
 In one-dimensional critical systems, the significant results achieved for quantum
 criticality include, for instance,
 the logarithmic finite-size scaling of the von Neumman entropy
 for bulk critical phenomena   \cite{Vidal,Calabrese,Korepin,Tagliacozzo,Pollmann}
 and the inverse finite-size correction of the geometric entanglement
 for boundary critical phenomena  \cite{Shi,Stephan,Hu,Liu,Affleck,Liu16}.
 Such characteristic features of critical systems
 have been shown that the scaling prefactors are respectively proportional to the central charge $c$
 and the correction factor $b$, which can be
 a fundamental quantity in conformal field theory and critical phenomena.

 However, entanglement is insufficient to describe the quantum character of
 correlations present in quantum states.
 Beyond entanglement, nontrivial quantum correlation can exist even in
 separable states, i.e.,
 entangled states are not the only kind of quantum states exhibiting nonclassical features.
 Actually, as a key concept in quantum information science, the quantum correlation among
 parts of a composite quantum system is a fundamental resource for several applications
 in quantum information.
 For a suitable measure of quantum correlations not only in entangled states but also
 in separable states, quantum discord has been introduced by Ollivier and Zurek \cite{Ollivier}.
 Quantum discord measures quantum correlations between two subsystems of a
 quantum many-body system \cite{Modi}.
 Similar to quantum entanglement,
 quantum discord has already been extensively studied in
 spin chain lattices at
 both zero  \cite{Dillenschneider,Sarandy,Maziero1} and
 finite \cite{Werlang,Werlang11} temperatures.
 In most cases,
 pairwise quantum discord between nearest-neighbor or next-nearest-neighbor spins
 has been considered and
 its singular behaviors has been focused to identify quantum phase transitions.
 Relatively less attention has been paid on the scaling of quantum discord \cite{Maziero12,Campell,Cakmak,Huang}.

 In principle,
 the origin of correlations can be either classical or quantum.
 In classical information theory,
 mutual information is a standard measure of correlation between two
 random variables.
 By replacing the Shannon entropy in the classical information theory
 with the von Neumann entropy,
 quantum mutual information can be defined \cite{Groisman, Adami, Alcaraz, Schumacher}
 and
 can quantify the total correlation, including
 both classical and quantum correlations, in a bipartite quantum state \cite{Groisman}.
 Compared to the conventional spatial correlations to characterize quantum phase transitions,
 quantum mutual information provides a more general
 tool to identify quantum phase transitions
 because one does not need to know a priori what the right correlation function is.
 Further,
 the measure for classical correlations can be defined as the maximal
 amount of information on one of the subsystems by operating a complete
 measurement process the other subsystem.
 Thus the definition of a proper measure of quantum correlations, i.e., quantum discord $\mathcal{D}$,
 is derived as
 the difference between the total correlations between two
 subsystems $A$ and $B$, represented by the quantum mutual
 information $\mathcal{I} (A : B)$, and the classical correlations $\mathcal{C}(A : B)$ \cite{Ollivier}.
 Similarly to the quantum discord,
 the mutual information and classical correlation can be used
 to identify quantum phase transitions.

 In this paper, our aim is then to find a relationship between
 conventional spatial correlation functions and quantum informational correlation functions.
 We investigate quantum mutual information, and classical and quantum
 correlation (quantum discord) in quantum critical systems.
 We consider the critical lines in the transverse-field spin-$1/2$ XY model.
 To obtain numerically the groundstate wavefunctions along the critical lines,
 the infinite matrix product state (iMPS) representation is employed
 with the infinite time-evolving block decimation (iTEBD) algorithm  \cite{Vidal03,Vidal07}.
 We will show that quantum mutual information, classical correlation and quantum discord can capture
 the characteristic features of criticality and obey a power-law decay with the increment of lattice distance in the critical regions for finite truncation dimensions.
 In particular, quantum mutual information, classical and quantum correlations
 are extrapolated to estimate their critical exponents
 in the thermodynamic limit.
 We find that with numerical errors, their critical exponents seem to be the same
 for each universality class.

 This paper is organized as follows.
 In Sec. II, we introduce the transverse-field spin-$1/2$ XY model that
 has the two characteristic transition lines.
 The exact critical exponents of the spin-spin correlation
 is mentioned from the exact analytic approach.
 A brief explanation is given for the numerical method, i.e.,
 the iMPS approach that is used in this study.
 We present the numerical procedure to obtain the critical exponents
 of the spin-spin correlation in the thermodynamic limit.
 The numerical critical exponents are shown to be well agreed with
 the corresponding exact values.
 In sec. III, the quantum mutual information is introduced as
 a measure of all kinds of correlations between two spins.
 Similar procedures are implemented to calculate and discuss the quantum mutual informations
 on the two transition lines.
 We estimate their critical exponents.
 As a part of the total correlation, the classical correlations are estimated in Sec. IV.
 Section V is devoted to discuss and calculate the quantum discord as a measure of quantum correlation.
 A summary and remarks of this work are given in Sec. VI.

\section{The transverse-field spin-$1/2$ XY spin chain}
 In investigating characteristic behaviors of classical and quantum correlations
 for quantum criticality,
 one of prototypical one-dimensional spin models can be
 the spin-$1/2$ quantum XY model given as the Hamiltonian,
\begin{equation}
 H = -\sum_{i=-\infty}^{\infty}
 \left[ \left(\frac{1+\gamma}{2}\right) \sigma^{x}_{i}\sigma^{x}_{i+1}
       + \left(\frac{1-\gamma}{2}\right) \sigma^{y}_{i}\sigma^{y}_{i+1}
       + h\, \sigma^{z}_{i} \right],
 \label{ham1}
\end{equation}
where $\sigma^{x,y,z}_i$ are the Pauli spin operators acting on $i$-th site,
$\gamma$ and $h$ are the anisotropy parameter
and the transverse magnetic field, respectively.
 This model is simply connected to the Ising model
 by taking the anisotropy parameter $\gamma=1$ or $\gamma=-1$,
 which means that for the anisotropy parameter $\gamma=1$ or $\gamma=-1$,
 the quantum critical points occur at $h=1$ or $h=-1$.
As is well-known, this XY model has two characteristic critical lines for the whole parameter ranges:
(i) the Ising transition lines are located at $\gamma \neq 0$ and $h = \pm 1 $, which separate
 the magnetically ordered and the disordered phases
 and (ii) the anisotropy transition line exists for $\gamma = 0 $ and $ -1 < h < 1$, which separate
 the magnetically ordered phases.
 These critical lines are distinguished in terms of the universality classes with central charges, i.e.,
 (i) the Ising university class with the central charge $c = 1/2$ and
 (ii) the Gaussian university class with $c = 1$, respectively.

\subsection{Exact critical exponents $\eta$ of the spin-spin correlations}

Actually, this transverse-field XY model can be exactly solved \cite{Lieb61,Katsura,Lieb66,Pfeuty}
by mapping the spins to spinless fermions via a Jordan-Wigner transformation \cite{Jordan}.
By using the exact analytic techniques,
Barouch and McCoy \cite{Barouch71} calculated
the correlation functions.
Further analysis on the correlation functions was made by
Damle and Sachdev \cite{Damle96}.
When  the spin-spin correlation function is defined as
 \begin{equation}
 \label{Classical}
 C_s(|i-j|)=\langle \sigma_{i}^{x}\, \sigma_{j}^{x} \rangle,
 \end{equation}
Bunder and McKenzie \cite{Bunder99} evaluated
the exact critical exponents $\eta_{ext}$ of the spin-spin correlations
as (i) $\eta_{ext} = 1/4$ on the Ising transition lines with $\gamma \neq 0$ and $h = \pm 1 $,
and (ii) $\eta_{ext} = 1/2$ one the anisotropy transition line
with $\gamma = 0 $ and $ -1 < h < 1$, respectively,
from the asymptotic form of the correlation function
in the limits of the infinite distance, i.e., $C_s^{asym}(|i-j|) \sim |i-j|^{-\eta}$.
The exact critical exponent of the spin-spin correlation on the Ising transition line in the transverse-field XY model was then shown to be consistent with the known exact exponent $\eta_{ext} = 1/4$ of the Ising model \cite{Lieb61,Pfeuty} with $\gamma=1$ or $\gamma=-1$. The two different values of the critical exponents
for the spin-spin correlation also characterize
the Ising transition lines and the anisotropy transition line, respectively.

 \subsection{\protect{i}MPS ground states and spin-spin correlations}

Each of the transition (critical) lines has the different physical
implications with the universality classes.
To numerically investigate characteristic behaviors of
classical and quantum correlations
for critical systems,
we then choose four different parameter sets in the two critical lines.
The four chosen parameter sets are (i) $(\gamma, h) = (1, 1) $ and $(1, 1/2)$
on the Ising transition line and (ii) $(\gamma, h) = (0, 0) $ and $(1/2, 0)$
on the anisotropy transition line.
 To consider the infinite size of the transverse-field spin-$1/2$ XY chain,
 we employ the infinite matrix product state (iMPS) representation \cite{Vidal03}.
 One can obtain ground state wavefunctions by using
 the infinite time evolving block decimation (iTEBD)
 algorithm \cite{Vidal07}.
 In principle, once an iMPS ground state wave function $\left|\psi\right\rangle$ is obtained,
 any type of conventional correlations including non-local correlations such as a string correlation
 can be calculated \cite{Vidal07,Su}.
 On employing this iMPS approach, previously several studies of various spin correlations have been performed,
 for instance, in spin-$1/2$ quantum Ising chain \cite{Vidal07}, ferromagnetic $4$-state \cite{Su}
 and antiferromagnetic $3$-state \cite{Dai17} Potts chains,
 quantum compass chain \cite{Wang},
 spin-$1$ XXZ chain \cite{Su12}, and so on.
 Especially, for the one-dimensional Ising model, the spin-spin correlation of the iMPS
 ground state wave function
 was demonstrated to decay as a power law at the critical point \cite{Vidal07}.

 To obtain iMPS ground state wavefunctions,
 we used first-order Trotter decomposition in the iTEBD algorithm  in this study.
 With an initial time step $dt = 0.1$, the time step is decreased according to a power law until $dt=10^{-6}$
 as the initial state approaches to a groundstate.
 Thus numerical iMPS wavefunctions are obtained for the truncation dimensions between $\chi = 20$
 and $\chi =150$.
 Before we investigate classical and quantum correlations,
 for a systematic study, we first consider the spin-spin correlations for the chosen parameters
 in our iMPS approach.

\begin{figure}
\includegraphics[width=0.5\textwidth]{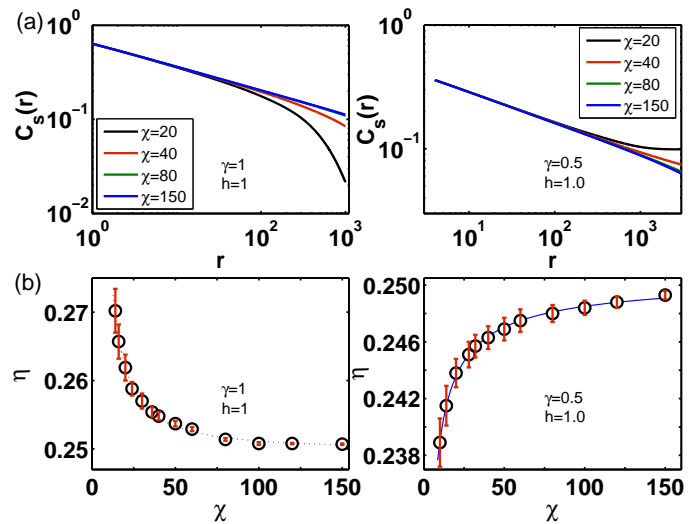}
\caption{(color online) (a) Spin-spin correlation $C_s(r)$ as
 a function of the lattice distance $r=|i-j|$ at
  $(\gamma,h)=(1.0,1.0)$ and
 $(0.5, 1.0)$ for various truncation dimension $\chi$.
 (b) Spin correlation exponent $\eta(\chi)$ as a function
 of truncation dimension $\chi$ at $(\gamma,h)=(1.0,1.0)$ and
 $(0.5,1.0)$.
 The exponent $\eta$ is given from the fitting function
 $C_s(r) = a_0 r^{-\eta}$ with the numerical constants $a_0$
 and $\eta$ for the power-law decaying part in (a).
 The details are discussed in the text.
 } \label{figure1}
\end{figure}

\subsubsection{Numerical critical exponent $\eta$ on the Ising transition line}
Let us first discuss spin-spin correlations on the Ising transition line.
In Fig.~\ref{figure1} (a), we plot the spin-spin correlation
at the parameters $(\gamma,h) = (1.0,1.0)$ and $(0.5,1.0)$
as a function of the lattice distance $r = |i-j|$
with the iMPS ground state wavefunctions for various truncation dimensions $\chi$.
For all truncation dimensions, the spin-spin correlations exhibits
a power-law decaying behavior.
The spin correlation value goes down to zero for larger distance.
For $(\gamma,h) = (1.0,1.0)$
the spin correlation functions show that the lattice range for a power-law decay becomes larger as the truncation dimension $\chi$ becomes larger.
The observed tendency of the spin correlation implies
that the power-law decaying range reaches an infinite lattice distance in the
thermodynamic limit if the truncation dimension $\chi \rightarrow \infty$.
In order to estimate the critical exponent $\eta$ for the spin-spin
correlation in the thermodynamic limit, we consider the
exponents of the power-law decaying part of the spin
correlation for the finite truncation dimensions. A numerical fit to the algebraically
decaying part is performed with the function
$C_s(r)=a_0 r^{-\eta}$ with the parameter values
(i) $a_0 = 0.650(3)$ and $\eta=0.262(2)$ for $\chi=20$,
(ii) $a_0 = 0.648(1)$ and $\eta = 0.2548(6)$ for $\chi=40$,
(iii) $a_0 = 0.6466(4)$ and $\eta = 0.2514(2)$ for $\chi=80$,
and
(iv) $a_0=0.6458(3)$ and $\eta=0.2507(1)$ for $\chi=150$.
The behavior of $\eta$ values show that the
exponent of $r$ in the fitting function appears to be approaching
the exact value $\eta_\infty = 1/4$ in the thermodynamic limit.
Thus,
we plot the estimates for
$\eta$ for finite truncation dimensions in Fig.~\ref{figure1} (b).
In order for the numerical estimation of the critical exponent $\eta_\infty$ in the thermodynamic limit,
we performed the extrapolation with the fitting function
$\eta(\chi) = \eta_0 \chi^a + \eta_\infty$,
The numerical constants are given as
$\eta_0 = 0.7(3)$, $a = -1.3(2)$, and $\eta_\infty = 0.2498(8)$.
This estimate $\eta_\infty$ at $(\gamma,h)=(1.0,1.0)$ is
in agreement with the exact value of the critical exponent $\eta_{ext}=1/4$
in the Ising model \cite{Lieb61,Pfeuty}.

In contrast to the case of the Ising model at $(\gamma,h) = (0.5,1.0)$,
for $(\gamma,h) = (0.5,1.0)$, the spin-spin correlation shows
a power-law decay to its saturated value.
The power-law decaying
part increases in distance from
a few hundreds to a few thousands of the lattice distance
as the truncation dimension increases.
Accordingly, the saturation value decreases.
One can expect that the power-law decaying range
becomes an infinite lattice distance in the
thermodynamic limit if the truncation dimension $\chi \rightarrow \infty$.
with the saturation value tending to zero.
In Fig. \ref{figure1}(a),
the exponents of the power-law decaying part of the spin
correlation are estimated with the function
$C_s(r)=a_0 r^{-\eta}$ with the parameter values
(i) $a_0 = 0.5037(5)$ and $\eta=0.2438(3)$ for $\chi=20$,
(ii) $a_0 = 0.506(1)$ and $\eta = 0.2463(8)$ for $\chi=40$,
(iii) $a_0 = 0.5077(9)$ and $\eta =0.2480(6)$ for $\chi=80$,
and
(iv) $a_0=0.5094(5)$ and $\eta=0.2493(3)$ for $\chi=150$.
The $\eta$ value increases to approach
the exact value $\eta_\infty = 1/4$ in the thermodynamic limit,
as the truncation dimension increases.
In Fig.~\ref{figure1} (b),
the estimates for $\eta$ are plotted for finite truncation dimensions.
In order for the numerical estimation of the critical exponent $\eta_\infty$ in the thermodynamic limit,
we performed the extrapolation with the fitting function
$\eta(\chi) = \eta_0 \chi^a + \eta_\infty$,
The numerical constants are given as
$\eta_0 = -0.060(9)$, $a = -0.71(8)$, and $\eta_\infty = 0.2508(6)$.
Similar to the case $(\gamma,h)=(1.0,1.0)$,
this estimate $\eta_\infty$ at $(\gamma,h)=(0.5,1.0)$ is
in agreement with the exact value of the critical exponent $\eta_{ext}=1/4$.

 %

\begin{figure}
\includegraphics[width=0.5\textwidth]{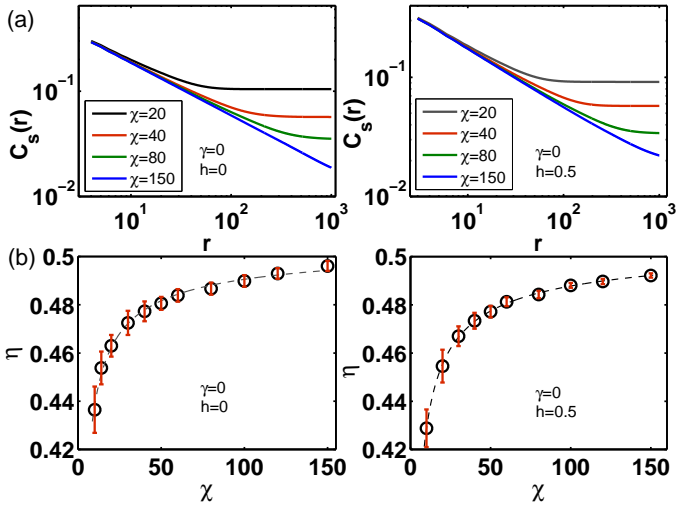}
\caption{(color online) (a) Spin-spin correlation $C_s(r)$ as
 a function of the lattice distance $r=|i-j|$ at
 $(\gamma,h)=(0,0)$ and
 $(0, 0.5)$ for various truncation dimension $\chi$.
 (b) Spin correlation exponent $\eta(\chi)$ as a function
 of truncation dimension $\chi$ at  $(\gamma,h)=(0,0)$ and
 $(0, 0.5)$.
 The exponent $\eta$ is given from the fitting function
 $C_s(r) = a_0 r^{-\eta}$ with the numerical constants $a_0$
 and $\eta$ for the power-law decaying part in (a).
 The details are discussed in the text.
 } \label{figure2}
\end{figure}

\subsubsection{Numerical critical exponent $\eta$ on the anisotropy transition line}
As was mentioned, the XY model has the other critical line, that is, the anisotropy transition line.
The critical line is situated at $\gamma=0$ for $-1 < h < 1$.
In Fig.~\ref{figure2} (a), we plot the spin-spin correlation
at the parameters $(\gamma,h) = (0.0,0.0)$ and $(0.0,0.5)$
as a function of the lattice distance $r = |i-j|$
with the iMPS ground state wavefunctions for various truncation dimensions $\chi$.
Similar to the case $(\gamma,h)=(0.5,1.0)$
on the Ising transition line in Fig. \ref{figure1} (a),
the spin-spin correlations in Fig. \ref{figure2} (a) show
a power-law decay to its saturated value for a given truncation dimension.
As the truncation dimension increases,
the power-law decaying
part increases in distance from
a few hundreds to a few thousands of the lattice distance
and the saturated value decreases.
Accordingly,
the power-law decaying range
may become an infinite lattice distance
with the saturation value tending to zero
in the thermodynamic limit.
With the function $C_s(r)=a_0 r^{-\eta}$,
we obtain the fitting constants as
for $(\gamma,h)=(0.0,0.0)$,
(i) $a_0 = 0.566(6)$ and $\eta=0.463(5)$ for $\chi=20$,
(ii) $a_0 = 0.572(5)$ and $\eta = 0.477(4)$ for $\chi=40$,
(iii) $a_0 = 0.576(4)$ and $\eta = 0.487(3)$ for $\chi=80$,
(iv) $a_0=0.582(2)$ and $\eta=0.496(2)$ for $\chi=150$.
Also in the case of
$(\gamma,h)=(0.0,0.5)$, the fitting constants are given as
(i) $a_0 = 0.524(7)$ and $\eta=0.455(7)$ for $\chi=20$,
(ii) $a_0 = 0.529(4)$ and $\eta = 0.473(3)$ for $\chi=40$,
(iii) $a_0 = 0.534(2)$ and $\eta = 0.484(2)$ for $\chi=80$,
and
(iv) $a_0=0.539(1)$ and $\eta=0.4922(9)$ for $\chi=150$.
These numerical fittings
show that as the truncation dimension increases,
the exponents of the power-law decaying part
become closer to $1/2$.
Then,
we numerically estimate
the spin correlation exponents in the thermodynamic limit
in Fig. (\ref{figure2}) (b).
To perform the extrapolation,
we employ the fitting function
$\eta(\chi) = \eta_0 \chi^a + \eta_\infty$.
The fitting results give the exponents $\eta_\infty=0.507(9)$
 with the constants $\eta_{0}=-0.28(9)$ and $a=-0.6(2)$
 at the critical point $(\gamma,h) = (0.0, 0.0)$,
 and
 $\eta_\infty=0.508(4)$ with $\eta_{0}=-0.31(3)$ and $a=-0.59(5)$
 at the critical point $(\gamma,h) = (0.0, 0.5)$.
 These results for both the critical exponents are consistent with the exact values $\eta_{ext} = 1/2$.

 Consequently,
 our iMPS approach shows
 that the critical exponents on both the Ising transition line and the anisotropy transition line are in excellent agreement with the known exact result $\eta_{ext}=1/4$ and $1/2$, respectively,
 for the spin-spin correlations.
 We adapt this same approach for the investigation of critical behaviors of various correlations
 defined in information science.

\section{Mutual information as a measure of all kinds of correlations between pairs of sites}
 In the previous section, we have numerically studied the traditional spin-spin correlation
 function between the two sites when the model system is in the critical systems.
 In this section, we consider a generalized correlation between two sites
 by employing entanglement-based measures
 in the perspective of information science.
 Since the two sites  distantly embedded in the infinite lattice system are in a mixed state,
 the origin of correlations between the two sites can be classical or quantum.
 Furthermore, there are quantum correlations which are not due to entanglement \cite{Dorner}.
 Thus in order to characterize all kinds of
 correlations between pairs of sites,
 we investigate the mutual information between the two sits.
 Actually, in classical information theory, mutual information is the standard measure of
 correlation between two arbitrary parts. Quantum mutual
 information can be defined as the quantum analog \cite{Groisman, Adami, Alcaraz, Schumacher}
 in terms of the von Neumann entropies, i.e.,
 the quantum mutual information between sites $i$ and $j$ is defined as
\begin{equation}
 {\mathcal{I}(i:j)=\mathcal{S}_{i}+\mathcal{S}_{j}-\mathcal{S}_{ij}},
 \label{Mutual}
\end{equation}
 where $\mathcal{S}_{i/i\cup j} =- \mathrm{Tr} \rho_{i/i\cup j} \log_2 \rho_{i/i\cup j}$
 are the von Neumann entropies
 with the reduced density matrix $\rho_{i/i\cup j}$ for one site $i$ and two sites $i\cup j$,
 respectively.
 In our iMPS approach, the full description of the groundstate of the quantum spin lattice model
 is given in a pure state by the iMPS wave function $|\psi\rangle$
 and thus the reduced density matrices $\rho_{i/i\cup j}$
 are
 obtained from the full density matrix $\rho = |\psi\rangle\langle\psi|$
 by tracing out the degrees of freedom of the rest of the
 subsystems $i^c$ or $(i\cup j)^c$, i.e.,
  $\rho_{i/i\cup j} = \mathrm{Tr}_{i^c/(i\cup j)^c} \, \rho$.
%

\begin{figure}
\includegraphics[width=0.5\textwidth]{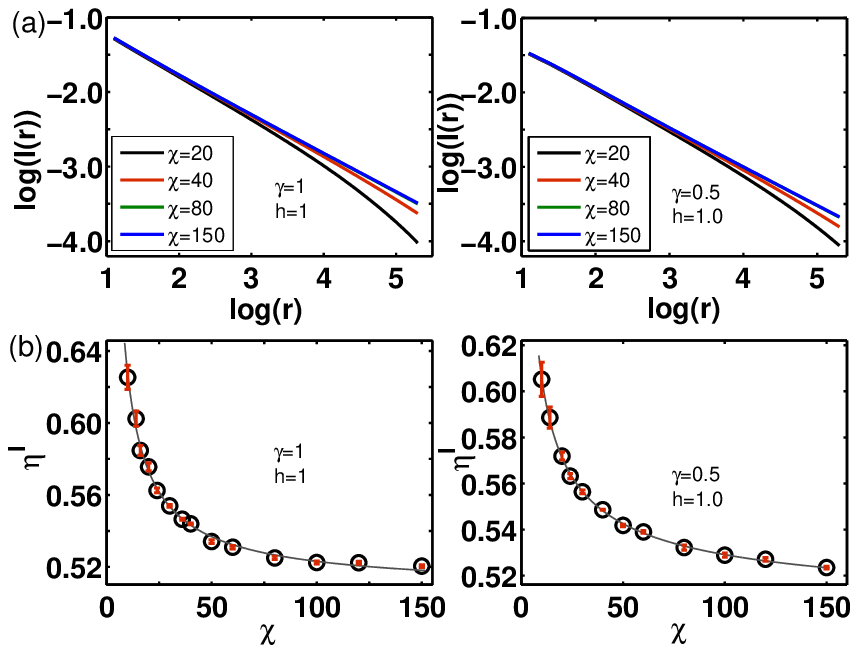}
\caption{(color online)
 (a) Mutual information ${\cal I}(r)$ as
 a function of the lattice distance $r=|i-j|$ at
  $(\gamma,h)=(1.0,1.0)$ and
 $(0.5, 1.0)$ for various truncation dimension $\chi$.
 (b) Mutual information exponent $\eta^I(\chi)$ as a function
 of truncation dimension $\chi$ at $(\gamma,h)=(1.0,1.0)$ and
 $(0.5,1.0)$.
 The exponent $\eta$ is given from the fitting function
  $\log {\cal I} (r)=-\eta^I \log \, r - a_0$  with the numerical constants $a_0$
 and $\eta^I$ for the power-law decaying part in (a).
 The details are discussed in the text.
 } \label{figure3}
\end{figure}

\subsection{Mutual information exponent $\eta^I$ on the Ising transition line}
 From our iMPS ground state wavefunctions for the chosen parameter,
 we obtain the necessary reduced density matrices for the quantum mutual information
 in Eq. (\ref{Mutual}).
 Let us first discuss quantum mutual information on the Ising transition line.
 In Fig.~\ref{figure3} (a), we plot the mutual information
 at the parameters $(\gamma,h) = (1.0,1.0)$ and $(0.5,1.0)$
 as a function of the lattice distance $r = |i-j|$
 for various truncation dimensions $\chi$.
 All the plots show that the mutual information decreases
 as the lattice distance $r$ increases.
 The log-log plots also show clearly that the decay of the mutual information
 seems to be linear.
 This implies that the mutual informations undergo a power-law decay to zero
 as the lattice distance increases to the infinity.
 Actually, the lattice range of the power-law decay becomes larger
 as the truncation dimension $\chi$ becomes larger
 and
 the slope of the mutual information in the log-log plots seem to be readily saturated for the
 truncation dimension $\chi = 150$.

 In order to confirm the power-law decay of the mutual information,
 we performed a numerical fit with the fitting function
 $\log {\cal I} (r)=-\eta^I \log \, r - a_0$ with the fitting constant $a_0$.
 For the critical point of the Ising model at $(\gamma,h) = (1.0,1.0)$,
 the detailed fitting constants are
(i) $a_0 = 0.647(6)$ and $\eta^I = 0.576(2)$ for $\chi=20$,
(ii) $a_0 = 0.690(1)$ and $\eta^I = 0.5439(4)$ for $\chi=40$,
(iii) $a_0 = 0.727(3)$ and $\eta^I = 0.5250(9)$ for $\chi=80$,
and
(iv) $a_0 = 0.740(3)$ and $\eta^I = 0.5204(7)$ for $\chi=150$.
 Similarly, for $(\gamma,h) = (0.5,1.0)$ on the Ising transition line,
 the numerical fittings are performed with the fitting constants:
(i) $a_0 = 0.813(5)$ and $\eta^I = 0.572(2)$ for $\chi=20$,
(ii) $a_0 = 0.848(3)$ and $\eta^I = 0.5486(8)$ for $\chi=40$,
(iii) $a_0 = 0.878(4)$ and $\eta^I = 0.5320(7)$ for $\chi=80$,
and
(iv) $a_0 = 0.905(3)$ and $\eta^I =0.5235(6)$ for $\chi=150$.
 This result implies that the  mutual information follows an asymptotic power-law scaling.
 For both the cases,
 the exponent of the power-law decay decreases as the truncation dimension increases.

 In Fig.~\ref{figure3} (b), we plot the estimates of $\eta^I$ for finite truncation dimensions.
 To obtain the exponent $\eta_{\infty}$ of the mutual information in the thermodynamic limit,
 we performed the extrapolation of $\eta^I$ with the numerical fitting function
 $\eta^I(\chi)=\eta^I_0 \chi^a+\eta_\infty$.
 The extrapolation reveals the mutual information exponent $\eta^I_\infty =0.506(7)$
 with  $\eta^I_0=0.8(2)$ and $a=-0.8(1)$ for $(\gamma,h) = (1.0,1.0)$,
 and $\eta^I_\infty =0.502(5)$
 with  $\eta^I_0=0.40(4)$ and $a=-0.59(6)$ for $(\gamma,h) = (0.5,1.0)$.
 Note that the exponents $\eta^I_\infty = 0.506(7)$ at $(\gamma,h) = (1.0,1.0)$
 and $\eta^I_\infty = 0.502(5)$ at $(\gamma,h) = (0.5,1.0)$ are very close to $1/2$.
 Consequently, similar to the spin-spin correlation,
 the mutual information on the Ising transition line
 undergoes a power-law decay to zero as the lattice distance increases,
 and its exponent of the power-law decay has a unique value, i.e., $\eta_\infty \simeq 1/2$.

\begin{figure}
\includegraphics[width=0.5\textwidth]{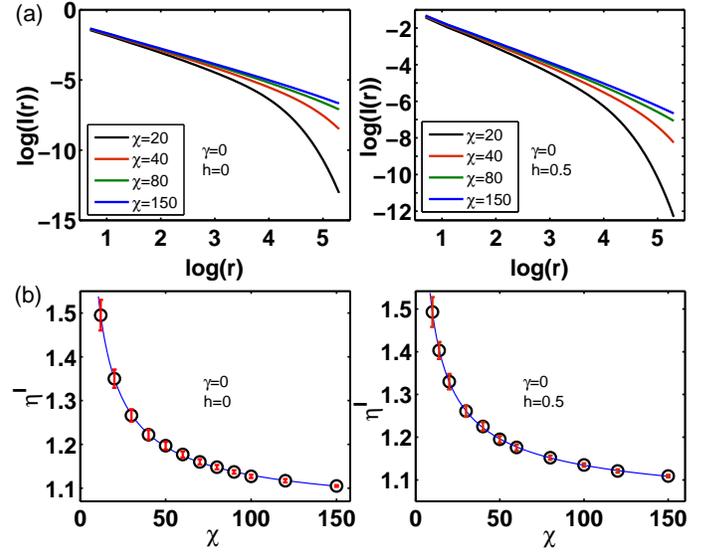}
\caption{(color online) (a) Mutual information ${\cal I}(r)$ as
 a function of the lattice distance $r=|i-j|$ at
 $(\gamma,h)=(0,0)$ and
 $(0, 0.5)$ for various truncation dimension $\chi$.
 (b) Mutual information exponent $\eta^I(\chi)$ as a function
 of truncation dimension $\chi$ at  $(\gamma,h)=(0,0)$ and
 $(0, 0.5)$.
 The exponent $\eta^I$ is given from the fitting function
 $\log {\cal I} (r)=-\eta^I \log \, r - a_0$  with the numerical constants $a_0$
 and $\eta^I$ for the power-law decaying part in (a).
 The details are discussed in the text.
 } \label{figure4}
\end{figure}

\subsection{Mutual information exponent $\eta^I$ on the anisotropy transition line}
 Next, let us discuss mutual information on the anisotropy transition line.
 Figure \ref{figure4} (a) displays the mutual informations
 at the parameters $(\gamma,h) = (0.0,0.0)$ and $(0.0,0.5)$
 as a function of the lattice distance $r = |i-j|$
 for various truncation dimensions $\chi$.
 As the lattice distance $r$ increases,
 the mutual information decreases in all the plots.
 Compared to the cases of the Ising transition line,
 the linear-decay range of the mutual information is shorter.
 However,
 the log-log plots show that the linear-decay range of the mutual information
 becomes larger as the truncation dimension increases.
 The slope of the mutual information in the log-log plots may be readily saturated for the
 truncation dimension $\chi = 150$.
 This implies that the mutual informations undergo a power-law decay to zero
 as the lattice distance increases to the infinity.

 The power-law decay of the mutual information on the anisotropy transition line
 can be shown with the numerical fits.
 The fitting function
 $\log {\cal I} (r)=-\eta^I \log \, r - a_0$
 is employed with the fitting constant $a_0$.
 For the XX model with the system parameter $(\gamma,h) = (0.0,0.0)$,
 the constants are fitted as
(i) $a_0 = 0.37(5)$ and $\eta^I = 1.35(2)$ for $\chi=20$,
(ii) $a_0 = 0.46(3)$ and $\eta^I = 1.22(1)$ for $\chi=40$,
(iii) $a_0 = 0.52(1)$ and $\eta^I = 1.148(5)$ for $\chi=80$,
and
(iv) $a_0 = 0.556(8)$ and $\eta^I = 1.105(2)$ for $\chi=150$.
 On the other parameter $(\gamma,h) = (0.0,0.5)$ in the anisotropy line,
 the numerical fittings are performed with the numerical constants:
(i) $a_0 = 0.41(4)$ and $\eta^I = 1.33(2)$ for $\chi=20$,
(ii) $a_0 = 0.46(3)$ and $\eta^I = 1.23(1)$ for $\chi=40$,
(iii) $a_0 = 0.52(1)$ and $\eta^I = 1.152(5)$ for $\chi=80$,
and
(iv) $a_0 = 0.56(1)$ and $\eta^I =1.109(3)$ for $\chi=150$.
 These results show that
 for a given truncation dimension,
 the exponents of the power-law decay give a very close value
 for the both parameters $(\gamma,h) = (0.0,0.0)$ and $(0.0,0.5)$

 In Fig.~\ref{figure4} (b), the estimates of $\eta^I$
 are plotted for finite truncation dimensions.
 The mutual information in the thermodynamic limit
 may give its exponent $\eta_{\infty}$ with the
 extrapolation function,
 $\eta^I(\chi)=\eta^I_0 \chi^a+\eta_\infty$.
 The extrapolation reveals the mutual information exponents $\eta^I_\infty =1.032(9)$
 with  $\eta^I_0=2.9(2)$ and $a=-0.73(3)$ for $(\gamma,h) = (0.0,0.0)$,
 and $\eta^I_\infty =1.015(9)$
 with  $\eta^I_0=1.91(8)$ and $a=-0.60(2)$ for $(\gamma,h) = (0.0,0.5)$.
 In contrast to the Ising transition line,
 these estimates show that the exponents $\eta^I_\infty = 1.032(9)$ at $(\gamma,h) = (0.0,0.0)$
 and $\eta^I_\infty =1.015(9)$ at $(\gamma,h) = (0.0,0.5)$ are very close to $1.0$.

 Consequently,
 the mutual information on both the Ising and the anisotropy transition lines
 follows  an asymptotic power-law scaling.
 Their critical exponents seems to be unique for each universality class.
 However, depending on the universality class,
 the values of the critical exponents of the mutual information
 are different, i.e., $\eta^I_\infty = 1/2$ for the Ising universality class
 and $\eta^I_\infty = 1$ for the Gaussian universality class.
 This result implies that the critical exponent of the quantum mutual information
 can be used to classify universality classes of critical systems.

\begin{figure}
\includegraphics[width=0.5\textwidth]{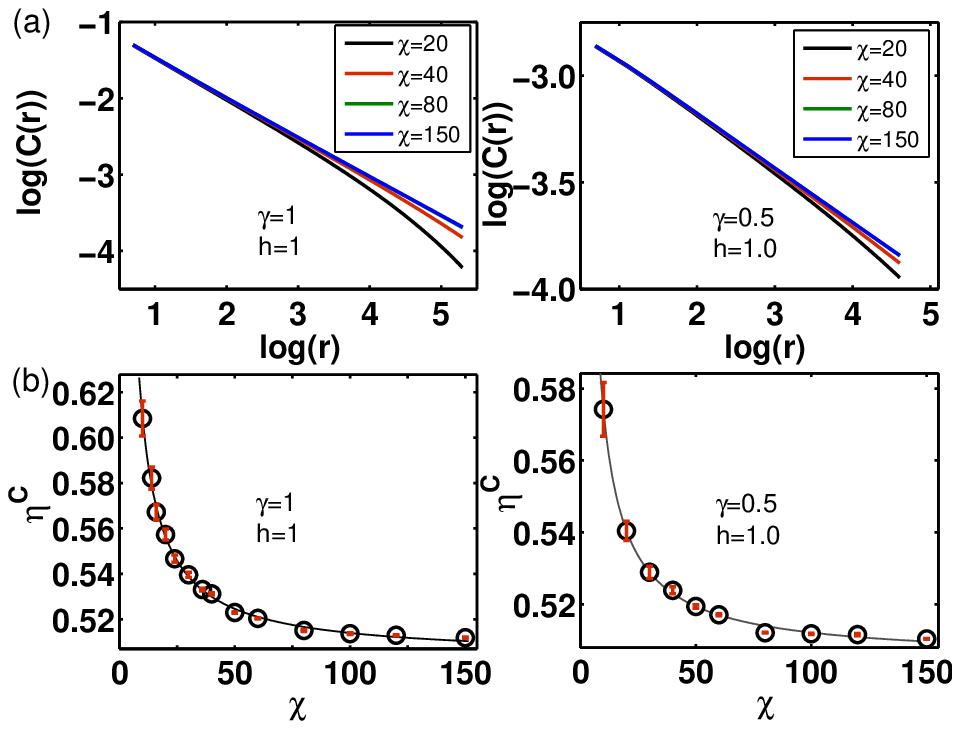}
\caption{(color online)  (a) Classical correlation ${\cal C}(r)$ as
 a function of the lattice distance $r=|i-j|$ at
  $(\gamma,h)=(1.0,1.0)$ and
 $(0.5, 1.0)$ for various truncation dimension $\chi$.
 (b) Classical correlation exponent $\eta^C(\chi)$ as a function
 of truncation dimension $\chi$ at $(\gamma,h)=(1.0,1.0)$ and
 $(0.5,1.0)$.
 The exponent $\eta$ is given from the fitting function
 $\log {\cal C} (r)=-\eta^C \log \, r - a_0$ with the numerical constants $a_0$
 and $\eta^C$ for the power-law decaying part in (a).
 The details are discussed in the text.
 } \label{figure5}
\end{figure}

\section{Classical correlation}
The quantum mutual information measures the total correlations within a quantum state,
which means that the total correlations contain both of classical and quantum correlations.
Splitting the total correlation (information) in a quantum state into a classical and a quantum parts
is possible by using a measurement on a part of the quantum state.
Actually, quantum mutual information conditioned on a complete set of von Neumann measurement
performed on one of subsystems corresponds to classical correlation part of the total correlation.
This classical correlation is defined as
\cite{Maziero1, Banerjee, Campbell, Li, He}
\begin{equation} \label{Classical}
 {\cal C}(\rho_{ij})={\cal S}(\rho_{i})-\min_{\Pi_{\alpha}}{\cal S}_{{\Pi_\alpha}}(\rho_{i|j})
\end{equation}
where the conditional entropy is defined as
\begin{equation}
 S_{{\Pi_\alpha}}(\rho_{i|j}) = \sum_{\alpha} q_{\alpha}S(\rho_{i}^{\alpha})
\end{equation}
with $q_{\alpha}=\mathrm{Tr}[(I_i\otimes\Pi_\alpha)\rho_{i\cup j}(I_i\otimes\Pi_\alpha)]$ and
$\rho_{i}^{\alpha}=(I_i\otimes\Pi_\alpha)\rho_{i\cup j}(I_i\otimes\Pi_\alpha)/q_\alpha$.
Here $I_i$ is the identity operator on the $i$-th site.
The minimum is taken over a complete set of projective measures $\Pi_\alpha$
on the partition $j$.
The complete set of orthonormal projectors onto the $j$-th site
 $ \Pi_\alpha=|\Theta_\alpha\rangle \langle\Theta_\alpha|$ with $\alpha \in \{ \parallel,\perp \}$,
where
$|\Theta_{\parallel}\rangle=\cos(\varphi/2)|0\rangle_{j}+e^{i\phi}\sin(\varphi/2)|1\rangle_{j}$,
and
$|\Theta_{\perp}\rangle= e^{-i\phi}\sin(\varphi/2)|0\rangle_{j}-\cos(\varphi/2)|1\rangle_{j}$.
This classical correlation is a measure to quantify the purely classical part of the total correlation
\cite{Henderson}.
By definition, the classical correlation is the maximum
amount of classical information that can be obtained about
one of the subsystems by performing local measurements on the
other of the subsystems.
Actually, much attentions have been paid on quantum correlation because
it plays a central role in
quantum information science, while
classical correlation does have relatively less attentions.
In this section,
as the counterpart of the quantum correlation,
classical correlation is evaluated in the critical systems.

\subsection{Classical correlation exponent $\eta^C$ on the Ising transition line}
 The classical correlation can be obtained by optimizing the $\varphi$ and $\phi$
 both in the regime $[0,\pi/2]$ to minimize ${\cal S}_{{\Pi_\alpha}}(\rho_{i|j})$ numerically
 in Eq. (\ref{Classical}).
 The classical correlation is considered first on the Ising transition line.
 From the iMPS groundstate wavefunctions for various truncation dimensions $\chi$,
 the classical correlations at the parameters $(\gamma,h) = (1.0,1.0)$ and $(0.5,1.0)$
 are calculated and displayed in Fig.~\ref{figure5} (a)
 as a function of the lattice distance $r = |i-j|$.
 As it should be, for all value of $r=|i-j|$,
 the overall amplitudes of the classical correlations are smaller than those of
 the quantum mutual information in Fig. \ref{figure3} (a).
 The log-log plots show clearly that the classical correlation
 seems to have a linear decay region.
 Actually, the lattice range for a power-law decay becomes larger
 as the truncation dimension $\chi$ becomes larger
 and
 the slope of the classical correlation in the log-log plots seems to be readily saturated for the
 truncation dimension $\chi = 150$.
 Similarly to the quantum mutual information,
 the classical correlations may undergo a power-law decay to zero
 as the lattice distance increases to the infinity.

 By introducing the same numerical fitting function
 $\log {\cal C} (r)=-\eta^C \log \, r - a_0$ with the fitting constant $a_0$,
 we study the power-law decay of the classical correlation.
 The Ising critical point at $(\gamma,h) = (1.0,1.0)$
 reveals the detailed fitting constants as follows;
(i) $a_0 = 0.908(6)$ and $\eta^C = 0.557(2)$ for $\chi=20$,
(ii) $a_0 = 0.939(2)$ and $\eta^C = 0.5312(5)$ for $\chi=40$,
(iii) $a_0 = 0.967(1)$ and $\eta^C = 0.5151(4)$ for $\chi=80$,
and
(iv) $a_0 = 0.975(1)$ and $\eta^C = 0.5119(3)$ for $\chi=150$.
 On the other chosen parameter $(\gamma,h) = (0.5,1.0)$
 in the Ising transition line, the numerical fittings are performed with
(i) $a_0 = 1.295(7)$ and $\eta^C = 0.540(3)$ for $\chi=20$,
(ii) $a_0 = 1.314(2)$ and $\eta^C = 0.524(1)$ for $\chi=40$,
(iii) $a_0 = 1.331(0)$ and $\eta^C = 0.5121(2)$ for $\chi=80$,
and
(iv) $a_0 = 1.335(1)$ and $\eta^C =0.5104(1)$ for $\chi=150$.
 As one can see,
 the exponent of the power-law decay region decreases as the truncation dimension increases
 for the both system parameters.

 The estimates of $\eta^C$ are plotted as a function of the finite truncation dimension
 in Fig.~\ref{figure5} (b).
 The exponent $\eta_{\infty}$ of the classical correlation in the thermodynamic limit
 can also be obtained with the numerical extrapolation function
 $\eta^C(\chi)=\eta^C_0 \chi^a+\eta^C_\infty$.
 From the extrapolation, the classical correlation exponents
 are estimated as $\eta^C_\infty = 0.503(4)$
 with  $\eta^C_0=1.1(2)$ and $a=-1.0(1)$ for $(\gamma,h) = (1.0,1.0)$,
 and $\eta^C_\infty =0.504(3)$
 with  $\eta^C_0=0.6(1)$ and $a=-1.0(1)$ for $(\gamma,h) = (0.5,1.0)$.
 One can notice clearly
 that the estimated exponents of the classical correlations, i.e.,
  $\eta^C_\infty =  0.503(4)$ at $(\gamma,h) = (1.0,1.0)$
 and $\eta^C_\infty = 0.504(3)$ at $(\gamma,h) = (0.5,1.0)$,
 may be considered to be the same with the exponents of the quantum mutual information in Subsec. III.A.
 Hence, similar to the quantum mutual information,
 the classical correlations
 follow the power-law scaling behavior
 and the critical exponent of the classical correlations is unique, i.e., $\eta^C_\infty = 1/2$
 in the Ising transition line.

\begin{figure}
\includegraphics[width=0.5\textwidth]{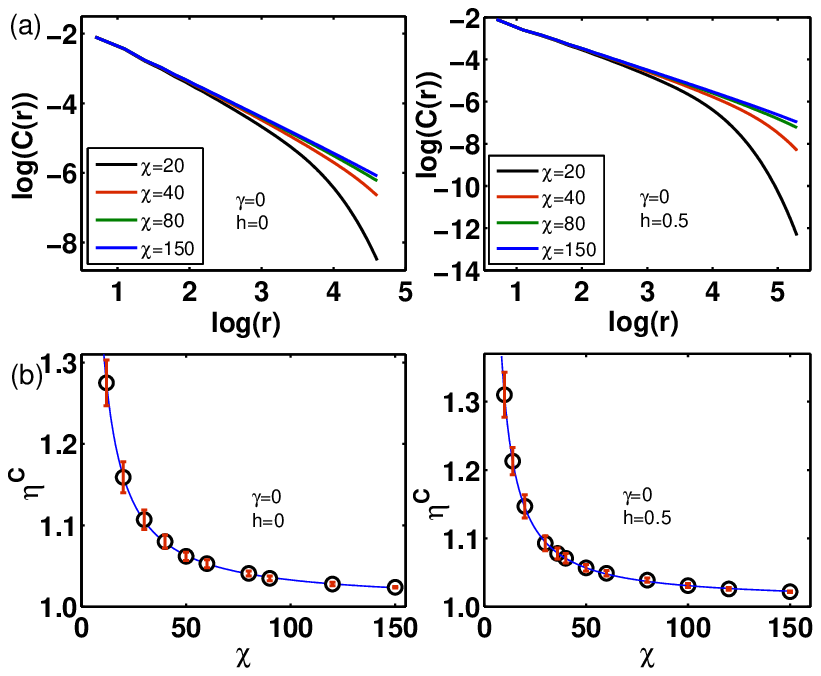}
\caption{(color online) (a) Classical correlation ${\cal C}(r)$ as
 a function of the lattice distance $r=|i-j|$ at
 $(\gamma,h)=(0,0)$ and
 $(0, 0.5)$ for various truncation dimension $\chi$.
 (b) Classical correlation exponent $\eta^C(\chi)$ as a function
 of truncation dimension $\chi$ at  $(\gamma,h)=(0,0)$ and
 $(0, 0.5)$.
 The exponent $\eta^C$ is given from the fitting function
 $\log {\cal C} (r)=-\eta^C \log \, r - a_0$ with the numerical constants $a_0$
 and $\eta^I$ for the power-law decaying part in (a).
 The details are discussed in the text.
 } \label{figure6}
\end{figure}

\subsection{Classical correlation exponent $\eta^C$ on the anisotropy transition line}
 The findings in the previous subsection is meaningful to understand the characteristic features
 of the correlations
 in critical systems.
 In order to see whether the findings are consistent,
 let us consider the classical correlation for the other universality class, i.e.,
  the anisotropy transition line.
 For the chosen parameters $(\gamma,h) = (0.0,0.0)$ and $(0.0,0.5)$ belonging to
 the Guassian universality class,
 the long-distance classical correlations are plotted in
 in Fig.~\ref{figure6} (a)
 as a function of the lattice distance $r = |i-j|$
 for various truncation dimensions $\chi$.
 For all value of $r=|i-j|$,
 the overall amplitudes of the classical correlations are smaller than those of
 the quantum mutual information in Fig. \ref{figure4} (a).
 All the plots exhibit a similar behavior with the classical correlations
 for the Ising universality class in Fig. \ref{figure5} (a).
 For power-law decay ranges of the classical correlations,
 the numerical fits have been performed with the fitting function
 $\log {\cal C} (r)=-\eta^C \log \, r - a_0$ with the constant $a_0$.
 For the XX model with the parameters $(\gamma,h) = (0.0,0.0)$,
 the detailed fitting constants are
(i) $a_0 = 1.14(4)$ and $\eta^C = 1.16(2)$ for $\chi=20$,
(ii) $a_0 = 1.24(2)$ and $\eta^C = 1.080(8)$ for $\chi=40$,
(iii) $a_0 = 1.30(1)$ and $\eta^C = 1.041(3)$ for $\chi=80$,
and
(iv) $a_0 = 1.328(5)$ and $\eta^C = 1.024(1)$ for $\chi=150$.
 For $(\gamma,h) = (0.0,0.5)$, the numerical fittings give
(i) $a_0 = 1.26(4)$ and $\eta^C = 1.15(2)$ for $\chi=20$,
(ii) $a_0 = 1.36(2)$ and $\eta^C = 1.071(7)$ for $\chi=40$,
(iii) $a_0 = 1.41(1)$ and $\eta^C = 1.039(3)$ for $\chi=80$,
and
(iv) $a_0 = 1.436(6)$ and $\eta^C =1.002(1)$ for $\chi=150$.
 In both cases, the exponent of the power-law decay decreases to approach $\eta^C = 1$ as the truncation dimension increases.

 In Fig.~\ref{figure6} (b), we plot the estimates of $\eta^C$ for finite truncation dimensions.
 To obtain the exponent $\eta_{\infty}$ of the classical correlation in the thermodynamic limit,
 we performed the extrapolation of $\eta^C$ with the numerical fitting function
 $\eta^C(\chi)=\eta^C_0 \chi^a+\eta^C_\infty$.
 The extrapolation reveals the classical correlation exponent $\eta^C_\infty =1.006(3)$
 with  $\eta^C_0=4.0(4)$ and $a=-1.08(4)$ for $(\gamma,h) = (0.0,0.0)$,
 and $\eta^C_\infty =1.009(2)$
 with  $\eta^C_0=4.2(3)$ and $a=-1.14(3)$ for $(\gamma,h) = (0.0,0.5)$.
 Similarly to the case of the Ising universality class,
 these exponents, i.e., $\eta^C_\infty = 1.006(3)$ at $(\gamma,h) = (0.0,0.0)$
 and $\eta^C_\infty = 1.009(2)$ at $(\gamma,h) = (0.0,0.5)$, means that
 the critical exponent of the classical correlation is $\eta^C_\infty  = 1$
 for the Gaussian universality class. These critical exponents
 are the same with those of the quantum mutual informations in Subsec. III.B.
 Together with the results for the Ising universality class,
 this fact implies that the critical quantum mutual information and critical classical correlation
 have the same exponent whose value is determined by universality class of critical systems.
 Consequently, quantum mutual information and classical correlation have the very characteristic feature of critical stems, that is, the power-law decaying behavior, and their critical exponent are
 unique to classify universality classes.

\section{Quantum correlation}
 So far we have demonstrated the critical features of the quantum mutual information and the classical correlation. We have also discussed that their critical exponents can classify
 universality classes of critical systems.
 As was discussed, the quantum mutual information ${\cal I}$ in Eq. (\ref{Mutual})
 measures the total correlation within a quantum state.
 The ${\cal C}$ in Eq. (\ref{Classical}) quantify all classical correlations.
 Once one obtains the quantum mutual information and the classical correlation,
 one can then obtain the nonclassical contributions to the total correlation by defining
 the difference between the total correlation and
 classical correlation \cite{Dillenschneider, Werlang, Tomasello, Banerjee, Lei, Campbell, Li, Maziero12,He}
\begin{equation}
 \mathcal{D}(\rho_{i\cup j})=\mathcal{I}(\rho_{i\cup j})-\mathcal{C}(\rho_{i\cup j}).
 \label{discord}
\end{equation}
Equation (\ref{discord}) is called quantum discord that is
zero for states with only classical correlation and nonzero
for states with quantum correlation.
Thus the quantum discord measures and quantifies all quantum correlations
including entanglement.

\begin{figure}
\includegraphics[width=0.5\textwidth]{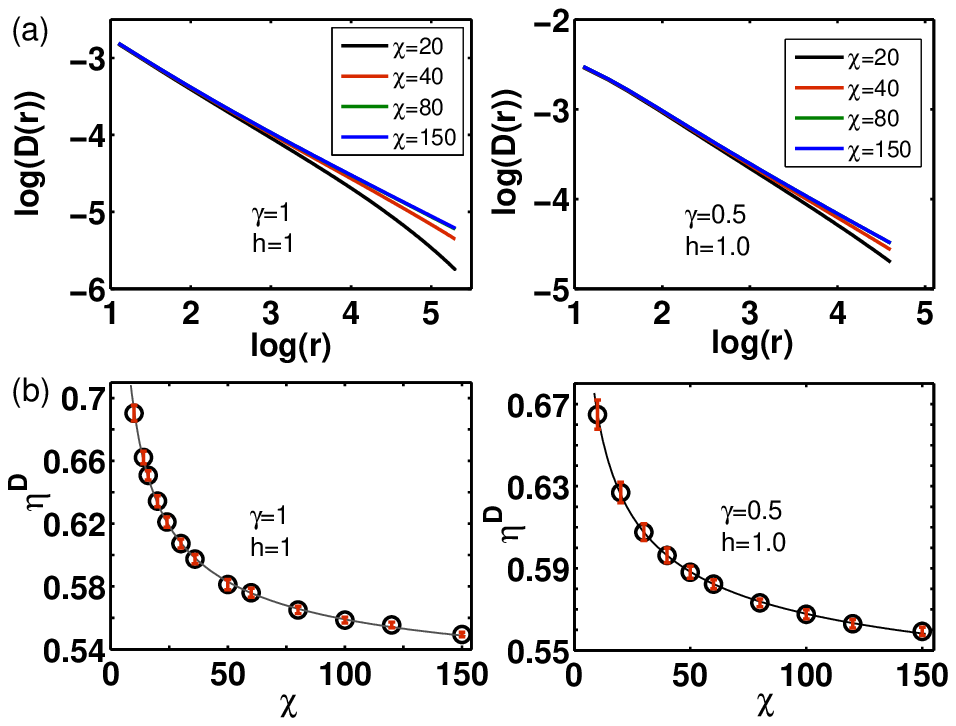}
\caption{(color online)  (a) Quantum correlation ${\cal D}(r)$ as
 a function of the lattice distance $r=|i-j|$ at
  $(\gamma,h)=(1.0,1.0)$ and
 $(0.5, 1.0)$ for various truncation dimension $\chi$.
 (b) Quantum correlation exponent $\eta^D(\chi)$ as a function
 of truncation dimension $\chi$ at $(\gamma,h)=(1.0,1.0)$ and
 $(0.5,1.0)$.
 The exponent $\eta$ is given from the fitting function
  $\log {\cal D} (r)=-\eta^D \log \, r - a_0$  with the numerical constants $a_0$
 and $\eta^D$ for the power-law decaying part in (a).
 The details are discussed in the text.
 } \label{figure7}
\end{figure}

\subsection{Quantum correlation exponent $\eta^D$ on the Ising transition line}

 Straightforwardly, within our iMPS calculation, the quantum correlation
 can be calculated from the quantum information and the classical correlation
 obtained in Sec. III and IV, respectively.
 To compared clearly with
 the quantum information in Sec. III and the classical correlation in Sec. IV,
 Let us first discuss quantum correlations on the Ising transition line.
 In Fig.~\ref{figure7} (a), the quantum correlations $\mathcal{D}(r)$
 are displayed for the parameters $(\gamma,h) = (1.0,1.0)$ and $(0.5,1.0)$
 as a function of the lattice distance $r = |i-j|$
 for various truncation dimensions $\chi$.
  As it should be,
 for all value of $r=|i-j|$,
 the overall amplitudes of the quantum correlations are smaller than those of
 the quantum mutual information in Fig. \ref{figure3} (a).
 Also, on can notice that
 the overall amplitudes of the quantum correlations are smaller than those of
 the classical correlation in Fig. \ref{figure5} (a).
 All the plots show that the quantum correlation decreases
 as the lattice distance $r$ increases.
 The tendency of quantum correlation is very similar to
 the quantum mutual information and classical correlation.
 The quantum correlation follows a power-law decay to zero
 as the truncation dimension increases.

 The power-law decay of the quantum correlation
 is fitted with the fitting function
 $\log {\cal D} (r)=-\eta^D \log \, r - a_0$.
 For the critical point of the Ising model $(\gamma,h) = (1.0,1.0)$,
 the detailed fitting constants are
 (i) $a_0 = 2.137(4)$ and $\eta^D = 0.634(4)$ for $\chi=20$,
 (ii) $a_0 = 2.204(9)$ and $\eta^D = 0.593(3)$ for $\chi=40$,
 (iii) $a_0 = 2.272(8)$ and $\eta^D = 0.565(2)$ for $\chi=80$,
 and
(iv) $a_0 = 2.322(6)$ and $\eta^D = 0.549(1)$ for $\chi=150$.
 Also, for $(\gamma,h) = (0.5,1.0)$, the numerical fittings are performed with
(i) $a_0 = 1.778(5)$ and $\eta^D = 0.627(5)$ for $\chi=20$,
(ii) $a_0 = 1.833(4)$ and $\eta^D = 0.596(4)$ for $\chi=40$,
(iii) $a_0 = 1.878(9)$ and $\eta^D = 0.573(2)$ for $\chi=80$,
 and
(iv) $a_0 = 1.925(7)$ and $\eta^D =0.559(2)$ for $\chi=150$.
 In Fig.~\ref{figure7} (b), we plot the estimates for $\eta^D$ for finite truncation dimensions.
 To obtain the exponent $\eta_{\infty}$ of the quantum correlation in the thermodynamic limit,
 we performed the extrapolation of $\eta^D$ with the numerical fitting function
 $\eta^C(\chi)=\eta^D_0 \chi^a+\eta^D_\infty$.
 The critical exponents of  quantum correlations
 are $\eta^D_\infty = 0.508(8)$
 with  $\eta^D_0=0.66(5)$ and $a=-0.56(5)$ for $(\gamma,h) = (1.0,1.0)$,
 and $\eta^D_\infty =0.505(9)$
 with  $\eta^D_0=0.41(2)$ and $a=-0.40(4)$ for $(\gamma,h) = (0.5,1.0)$.
 Both the estimates $\eta^D_\infty =  0.508(8)$ at $(\gamma,h) = (1.0,1.0)$
 and $\eta^D_\infty = 0.505(9)$ at $(\gamma,h) = (0.5,1.0)$ reveal
 a unique critical exponent of quantum correlation, i.e.,
 $\eta^D_\infty = 1/2$ in the Ising transition line.
 For the Ising model at $(\gamma,h) = (1.0,1.0)$,
 our result is consistent with the analytical value $\eta^D_\infty=1/2$
 from the asymptotic form of the quantum correlation function \cite{Huang}.
 Interestingly, this fact suggests that
 all the critical exponents of quantum mutual information and
 the classical and quantum correlation
 exhibit the same critical features that are the power-law decay to zero as the lattice distance increases
 and their unique critical exponents are the same value
  $\eta^I_\infty = \eta^C_\infty =\eta^D_\infty = 1/2$
 for the Ising universality class.

\begin{figure}
\includegraphics[width=0.5\textwidth]{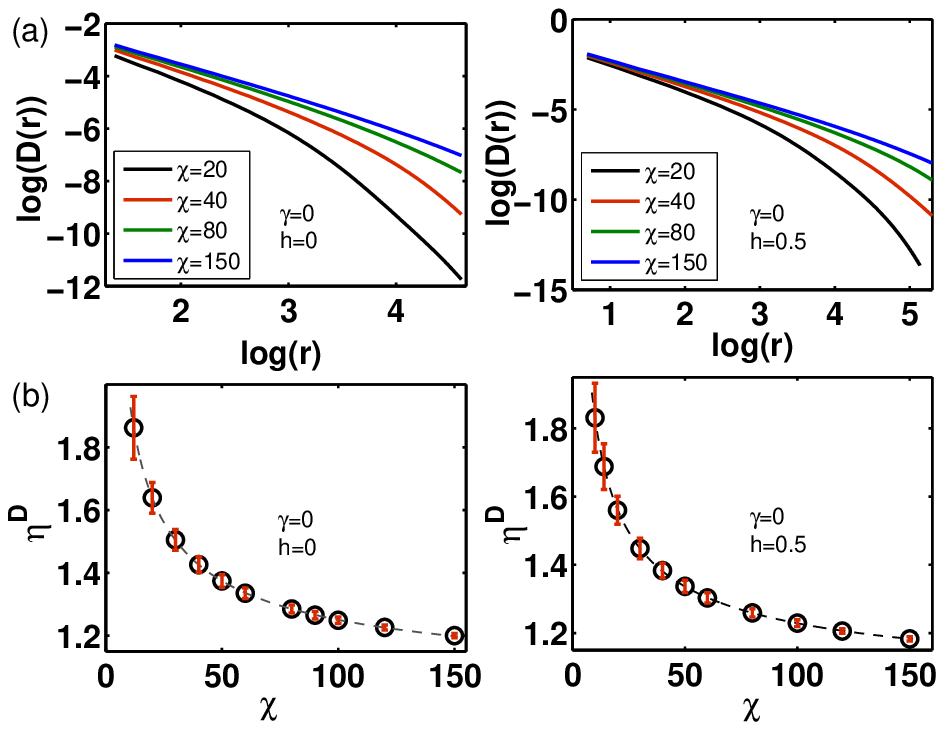}
\caption{(color online) (a) Quantum correlation ${\cal D}(r)$ as
 a function of the lattice distance $r=|i-j|$ at
 $(\gamma,h)=(0,0)$ and
 $(0, 0.5)$ for various truncation dimension $\chi$.
 (b) Quantum correlation exponent $\eta^D(\chi)$ as a function
 of truncation dimension $\chi$ at  $(\gamma,h)=(0,0)$ and
 $(0, 0.5)$.
 The exponent $\eta^D$ is given from the fitting function
 $\log {\cal D} (r)=-\eta^D \log \, r - a_0$  with the numerical constants $a_0$
 and $\eta^D$ for the power-law decaying part in (a).
 The details are discussed in the text.
 } \label{figure8}
\end{figure}

\subsection{Quantum correlation exponent $\eta^D$ on the anisotropy transition line}
 To show that the critical features in
 all correlations, i.e., quantum mutual information and
 classical and quantum correlation, hold for the Gaussian universality class
 in the transverse-field spin-$1/2$ XY model,
 let us discuss quantum correlations on the anisotropy transition line.
 Similarly to the cases of the previous section,
 we plot the quantum correlation
 at the parameters $(\gamma,h) = (0.0,0.0)$ and $(0.0,0.5)$
 as a function of the lattice distance $r = |i-j|$ in Fig.~\ref{figure8} (a).
 Comparing with Figs. \ref{figure4} (a) and \ref{figure6} (a),
 Fig ~\ref{figure8} (a) confirms that
 the overall amplitudes of the quantum correlations are smaller than those of
 the quantum mutual informations
 and the classical correlations.
 All the log-log plots also show that the lattice range for a power-law decay becomes larger
 as the truncation dimension $\chi$ becomes larger.
 This fact can be shown by performing the numerical fitting on the power-law decays of the quantum correlations with the fitting function
 $\log {\cal D} (r)=-\eta^D \log \, r - a_0$.
 For the critical point $(\gamma,h) = (0.0,0.0)$,
 the detailed fitting constants are
(i) $a_0 = 0.94(9)$ and $\eta^D = 1.64(5)$ for $\chi=20$,
(ii) $a_0 = 1.01(6)$ and $\eta^D = 1.43(3)$ for $\chi=40$,
(iii) $a_0 = 1.09(4)$ and $\eta^D = 1.29(1)$ for $\chi=80$,
and
(iv) $a_0 = 1.15(2)$ and $\eta^D = 1.200(7)$ for $\chi=150$.
 Also, for $(\gamma,h) = (0.0,0.5)$, the numerical fittings are performed with
(i) $a_0 = 0.94(8)$ and $\eta^D = 1.56(4)$ for $\chi=20$,
(ii) $a_0 = 0.97(5)$ and $\eta^D = 1.38(2)$ for $\chi=40$,
(iii) $a_0 = 1.05(3)$ and $\eta^D = 1.26(1)$ for $\chi=80$,
and
(iv) $a_0 = 1.10(2)$ and $\eta^D =1.183(6)$ for $\chi=150$.
 For both cases, the exponent of the power-law decay decreases as the truncation dimension increases,
 as is shown in Fig. ~\ref{figure8} (b).

 Figure~\ref{figure8} (b) shows the estimates for $\eta^D$ for finite truncation dimensions.
 The critical exponents of the quantum correlations $\eta^D_{\infty}$ in the thermodynamic limit,
 are obtained with the extrapolation function
 $\eta^D(\chi)=\eta^D_0 \chi^a+\eta_\infty$.
 The extrapolation give the exponents $\eta^D_\infty =1.008(8)$
 with  $\eta^D_0=3.74(8)$ and $a=-0.59(1)$ for $(\gamma,h) = (0.0,0.0)$,
 and $\eta^D_\infty =1.005(8)$
 with  $\eta^D_0=3.05(6)$ and $a=-0.57(1)$ for $(\gamma,h) = (0.0,0.5)$.
 For the XX model at $(\gamma,h) = (0.0,0.0)$,
 our result is consistent with the analytical value $\eta^D_\infty=1$
 from the asymptotic form of the quantum correlation function \cite{Huang}.
 Hence,
 for the anisotropy transition line,
 our estimates $\eta^D_\infty \simeq 1.0$
 offers that
 all the critical exponents of quantum mutual information and
 the classical and quantum correlation
 exhibit  the unique value
  $\eta^I_\infty = \eta^C_\infty =\eta^D_\infty = 1$
 for the Gaussian universality class.

\begin{table*}
\renewcommand\arraystretch{2}
\caption{Comparisons between the numerical and the numerical critical exponents
 for the Ising and the anisotropy transition lines  in the transverse-field spin-$1/2$ XY model
the critical regions}
\begin{tabular}{c|c|c|c|c|c|c|c}
\hline
\hline
      & \begin{minipage}{1.5cm} Mutual \\ information \end{minipage} &
      \begin{minipage}{1.5cm} Classical \\ correlation \end{minipage} &
      \begin{minipage}{1.5cm} Quantum \\ discord \end{minipage} &
      \begin{minipage}{1.5cm} Spin-spin \\ correlation \end{minipage} &
      \begin{minipage}{1.5cm} Quantum \\ discord \end{minipage} &
      \begin{minipage}{1.5cm} Spin-spin \\ correlation \end{minipage} &
      \begin{minipage}{1.5cm} Central \\ charge \end{minipage} \\
 & $\eta_{num}^{I}$ & $\eta_{num}^{C}$ & $\eta_{num}^{D}$ & $\eta_{num}$ & $\eta_{exa}^{D}$ & $\eta_{exa}$ & $c$ \\
\hline \hline
  \begin{minipage}{2.5cm} Ising \\ $(\gamma=1.0, h=1.0)$  \end{minipage}
 & 0.506(7) & 0.503(4) & 0.508(8) & 0.2498(8) & 1/2 \cite{Huang} & 1/4 \cite{Bunder99} &  1/2\\
\hline
 \begin{minipage}{2.5cm} XY \\ $(\gamma=0.5, h =0.5)$ \end{minipage}
 & 0.502(5) & 0.504(3) & 0.505(9) & 0.2508(6) & 1/2 & 1/4 \cite{Bunder99} & 1/2\\
\hline
 \begin{minipage}{2.5cm} XX \\ $(\gamma=0.0, h = 0.0)$ \end{minipage}
 & 1.032(9) & 1.006(3) & 1.008(8) & 0.507(9) & 1 \cite{Huang} & 1/2 \cite{Bunder99} & 1\\
\hline 
 \begin{minipage}{2.5cm} XY\\  $(\gamma=0.0, h=0.5)$ \end{minipage}
 & 1.015(9) & 1.009(2) & 1.005(8) & 0.508(4) &  & 1/2 \cite{Bunder99}& 1 \\
 \hline
 \hline
\end{tabular}
\label{table}
\end{table*}

\section{Summary}

 The classical and quantum correlations defined from quantum information science have been
 numerically investigated in one-dimensional quantum spin-$1/2$ lattice systems.
 In order to study quantum criticality with respect to the classical and quantum correlations,
 we have considered the infinite-size spin chain by employing
 the iMPS representation with the ITEBD algorithm.
 For the transverse-field spin-$1/2$ XY model,
 we calculated the iMPS ground state wavefunctions
 in the two characteristic critical lines,
 i.e., the Ising and the anisotropy transition lines.
 Calculated from our iMPS ground state wavefunctions,
 the traditional spin-spin correlations are shown to exhibit the power-law decays
 at the chosen critical points.
 By using the extrapolation of the exponents of the spin-spin correlations
 for finite truncation dimensions,
 the critical exponents $\eta$ are estimated in excellent agreements with the known values from the exact solution.

 The same approach was adapted to investigate the critical behaviors of the classical and the quantum correlations as well as the quantum mutual information.
 All of the correlations are found to exhibit a consistent power-law decaying behavior for various truncation dimensions.
 The estimated exponents of the power-law decaying regions
 were extrapolated to obtain the critical exponents
 of the correlations in the thermodynamic limit.
 In Table \ref{table}, we summarize the numerical estimates
 of the critical exponents with the known exact values
 of some critical exponents.
 The comparison in Table \ref{table} shows clearly that
 the quantum mutual information and the classical and the quantum correlations
 have almost same values, i.e., $\eta^I \simeq \eta^C \simeq \eta^D$
 for both the Ising and the Gaussian universality classes.
 The critical exponents were estimated as
 $\eta^I \simeq \eta^C \simeq \eta^D \simeq 1/2$ and $1$ for the Ising and the anisotropy transition lines,
 respectively.
 It is also shown that regardless of the universality classes,
 the critical exponents of the correlations defined from quantum information science
 has a universal relation with those of the spin-spin correlations, i.e.,
 $\eta = \eta^\alpha/2$.
 Therefore, our results suggest that
 the quantum mutual information and the classical and the quantum correlations
 can capture a characteristic feature of quantum critical systems
 and
 their critical exponents can be used to
 characterize the universality class of quantum critical systems.

{\it Acknowledgements.}
 SYC acknowledges support in part from the
 National Natural Science Foundation of China (Grants No. 11674042,
 and No. 11174375).
 Y.W.D. and D.X.Y. are supported by NKRDPC-2017YFA0206203, NSFC-11574404,
 NSFG-2015A030313176, the National Supercomputer Center in Guangzhou,
 and the Leading Talent Program of Guangdong Special Projects.


\end{document}